\documentclass[%
 reprint,
 amsmath,
 amssymb,
 aps,
 prd,
 nofootinbib,
 superscriptaddress,
 showkeys
]{revtex4-2}
\usepackage{ulem}
\usepackage{graphicx}
\usepackage{xcolor}
\usepackage{mathtools}
\usepackage{tabularray}
\UseTblrLibrary{booktabs}
\graphicspath{{./}{figures/}}
\usepackage{verbatim} 
\usepackage{bm}
\usepackage{hyperref}
\usepackage{orcidlink}
\hypersetup{
    colorlinks=true,
    citecolor=blue,
    urlcolor=blue,
}

\begin{document}

\title{Dark matter coupled to radiation: Limits from the Milky Way satellites}

\author{Wendy Crumrine\orcidlink{0009-0002-7428-7019}}
\email{crumrine@usc.edu}
\affiliation{Department of Physics $\&$ Astronomy, University of Southern California, Los Angeles, California, 90007, USA}

\author{Ethan O.~Nadler\orcidlink{0000-0002-1182-3825}}
\email{enadler@carnegiescience.edu}
\affiliation{Carnegie Observatories, 813 Santa Barbara Street, Pasadena, California 91101, USA}
\affiliation{Department of Physics $\&$ Astronomy, University of Southern California, Los Angeles, California, 90007, USA}
\affiliation{Department of Astronomy \& Astrophysics, University of California, San Diego, La Jolla, California 92093, USA}

\author{Rui An\orcidlink{0000-0001-9543-5012}}
\email{anrui@usc.edu}
\affiliation{Department of Physics $\&$ Astronomy, University of Southern California, Los Angeles, California, 90007, USA}

\author{Vera Gluscevic\orcidlink{0000-0002-3589-8637}}
\email{vera.gluscevic@usc.edu}
\affiliation{Department of Physics $\&$ Astronomy, University of Southern California, Los Angeles, California, 90007, USA}

\begin{abstract}
Interactions between dark matter (DM) and relativistic particles at early times suppress structure formation on small scales. In particular, the scattering process transfers heat and momentum from radiation to DM, ultimately reducing the abundance of low-mass DM halos and the dwarf galaxies they host. Herein, we derive limits on DM--photon and DM--neutrino scattering cross section using the Milky Way satellite galaxy population. We consider temperature-independent interactions parameterized by DM mass ($m_\chi$) and DM--radiation interaction cross section ($\sigma_{\chi\text{--}i}$, where $i$ represents the target species). By requiring that the linear matter power spectra be strictly less suppressed than in the case of a thermal-relic warm DM cutoff, we derive the following $95\%$ upper limits at $m_\chi=1$ MeV: $\sigma_{\chi\text{--}\gamma}<1.98\times10^{-38}\text{cm}^2$ and $\sigma_{\chi\text{--}\nu}<3.16\times10^{-38}\text{cm}^2$. Our bounds on $\sigma_{\chi\text{--}i}$ depend linearly on $m_\chi$ for $m_\chi \gtrsim 1~\mathrm{MeV}$ and improve upon previous limits by 1 order of magnitude. The mass dependence of our limit approaches $m_\chi^3$ at lower masses due to the effects of DM sound speed; at $m_{\chi}=100~\mathrm{keV}$, we arrive at an upper limit 3 orders of magnitude more stringent than achieved in previous explorations. Upcoming dwarf galaxy surveys will further improve the sensitivity of similar analyses, complementing laboratory and indirect detection searches for DM--radiation interactions.

\end{abstract}

\maketitle

\section{Introduction}\label{sec:intro}

Diverse astrophysical probes suggest that roughly one quarter of our Universe's energy budget consists of dark matter (DM), i.e.,\ matter beyond the Standard Model (SM) of particle physics. Thus far, the particle nature of DM remains a mystery, while evidence for its existence arises solely from astrophysical observations. While laboratory experiments have historically provided stringent constraints on DM properties, cosmology has emerged as a complementary test of its microphysics.  

The standard paradigm of cold, collisionless dark matter (CDM), which interacts only through gravity, agrees with the large-scale structure of the Universe, as probed by the cosmic microwave background radiation (CMB), galaxy clustering, and Lyman-$\alpha$ forest measurements \cite{Chabanier_2019}. More recent work shows that CDM is broadly consistent with halo abundances on small, nonlinear scales, as probed by dwarf galaxy populations \cite{Nadler_2020,Jethwa_2017}. Other observations promise to further test consistency at small scales using, e.g., strong gravitational lensing \cite{Hsueh_2019, Gilman_2019,Keeley:2024brx}, stellar streams \cite{Banik_2018, Bonaca_2019}, and dwarf galaxy stellar velocity dispersions \cite{Kim_2021,Esteban:2023xpk}. However, observational incompleteness and theoretical uncertainties surrounding halos smaller than $\sim$$10^8~M_{\mathrm{\odot}}$ provide space for beyond-CDM models as viable avenues for investigation. 

Such nonstandard models can result in suppression of power on scales probed by small halos, for a variety of physical reasons: DM free streaming (in the case of warm DM, WDM; \cite{1988ApJ...332....1S,Sommer_Larsen_2001}), a sizeable DM de-Broglie wavelength (fuzzy DM, FDM; \cite{Hu_2000}), heat and momentum exchange due to interactions with SM particles or dark radiation (interacting DM, IDM; \cite{Boehm_2005,Boddy_2018,Census_III,Nadler_2019_b,Maamari_2021,zhou2024searchingdarkmatterinteractions,Vogelsberger_2016}). In each of these scenarios, linear density perturbations on small scales are suppressed, reducing the abundance of low-mass DM halos throughout cosmic history, and the dwarf galaxies they host. We herein explore a particular class of IDM models; namely, DM that exhibits interactions with SM radiation prior to the onset of structure formation.

We consider a cosmology in which 100$\%$ of DM scatters elastically with either photons ($\gamma$--DM) or neutrinos ($\nu$--DM) via an effective momentum-transfer cross section $\sigma_{\chi\text{--}i}$ (where $i$ represents the target species) that is constant with temperature. Such interactions can arise in various particle physics scenarios, including a millicharge coupling with Thomson scattering-like properties (for $\gamma$--DM) or a DM coupling to a sterile neutrino (for $\nu$--DM; \cite{Brax_2023_1,Brax_2023,Paul_2021}). Previous analyses have constrained temperature-independent DM--radiation interactions using CMB anisotropies \cite{Wilkinson_2014_nu,Wilkinson_2014_photon,Escudero_2015,Stadler_2018,Paul_2021}, Lyman-$\alpha$ forest flux power spectra \cite{Hooper_2022}, the Milky Way (MW) satellite galaxy population \cite{B_hm_2001,B_hm_2014,Escudero_2018,Akita_2023}, the cosmic reionization history \cite{Dey_2023}, high-energy astrophysical events \cite{Ferrer_2023,Fujiwara_2024,Cline_2023,Cline_2023_b}, dwarf galaxy density profiles \cite{ heston2024constraining}, and combinations of datasets \cite{Becker_2021,Brax_2023,Brax_2023_1,giarè2023hints}. 

In this work, we derive new leading limits on DM--radiation interactions using the latest census of the MW satellite galaxies, measured by the Dark Energy Survey (DES) and Pan-STARRS1 (PS1); see Ref.~\cite{Drlica_Wagner_2020}. By leveraging the bound on  thermal-relic WDM mass from Ref.~\cite{Census_III} ($m_{\mathrm{WDM}}>6.5$ $\text{keV}$ at $95\%$ confidence), we obtain upper limits on $\sigma_{\chi\text{--}\gamma}$ and $\sigma_{\chi\text{--}\gamma}$. These limits exclude models that suppress power at scales corresponding to the halos that host the faintest observed dwarf galaxies, i.e., halos of mass $M$ $\sim$ $10^8$$M_{\mathrm{\odot}}$, corresponding to wavenumbers $k$ $\sim$ $10$-$100$ $h~\mathrm{Mpc}^{-1}$ in linear theory (see Eq.\ 5 from Ref.~\cite{Nadler_2019_b}).
 
Our analysis  goes beyond previous literature by including the effects of DM sound speed, which has a non-negligible impact on the linear matter power spectrum, $P(k)$, at low DM masses. Previous investigations considering these scenarios have excluded the effects of DM sound speed, either due to approximations in numerical treatments or, in the case of CMB analyses, negligible effects at the scales probed \cite{Stadler_2018}. The mass dependence of the $\sigma_{\chi\text{--}i}$ bound, which is linear in the absence of DM sound speed, transitions toward a steeper scaling for $m_\chi\lesssim1$ MeV, approaching $m_\chi^3$ at lower masses. Thus, the inclusion of DM sound speed improves the $100$ keV bound---for both DM--neutrino and DM--photon scattering---by 2 orders of magnitude, and results in an upper limit 3 orders of magnitude more constraining than previous limits set using the MW satellites \cite{Escudero_2018, Akita_2023}. For DM masses larger than $1$ MeV, our bound improves upon these satellite bounds by 1 order of magnitude. We derive these constraints assuming massless neutrinos, which we show is a good approximation for the observables considered.

This paper is structured as follows: In Section \ref{sec:theory}, we provide theoretical background on DM--photon and DM--neutrino scattering scenarios and discuss their effects on the linear matter power spectrum. In Section \ref{sec:approach}, we summarize our approach for deriving constraints on interaction strength for these scenarios, and we present our results in Section \ref{sec:results}. We discuss our results and conclude in \ref{sec:conclusion}. Throughout, we adopt the following cosmological parameters, following Ref.~\cite{Maamari_2021}: Hubble constant $h=0.6932$, baryon density $\Omega_bh^2=0.02223$, DM density $\Omega_{\mathrm{dm}}h^2=0.1153$, optical depth to reionization $\tau_{\mathrm{reio}}=0.081$, scalar perturbations amplitude $A_s=2.464\times10^{-9}$, scalar spectral index $n_s=0.9608$, and effective number of neutrino species $N_{\mathrm{eff}}=3.046$.

\section{Theory}\label{sec:theory}

\begin{figure*}[htbp]
    \centering
    \includegraphics[width=\linewidth]{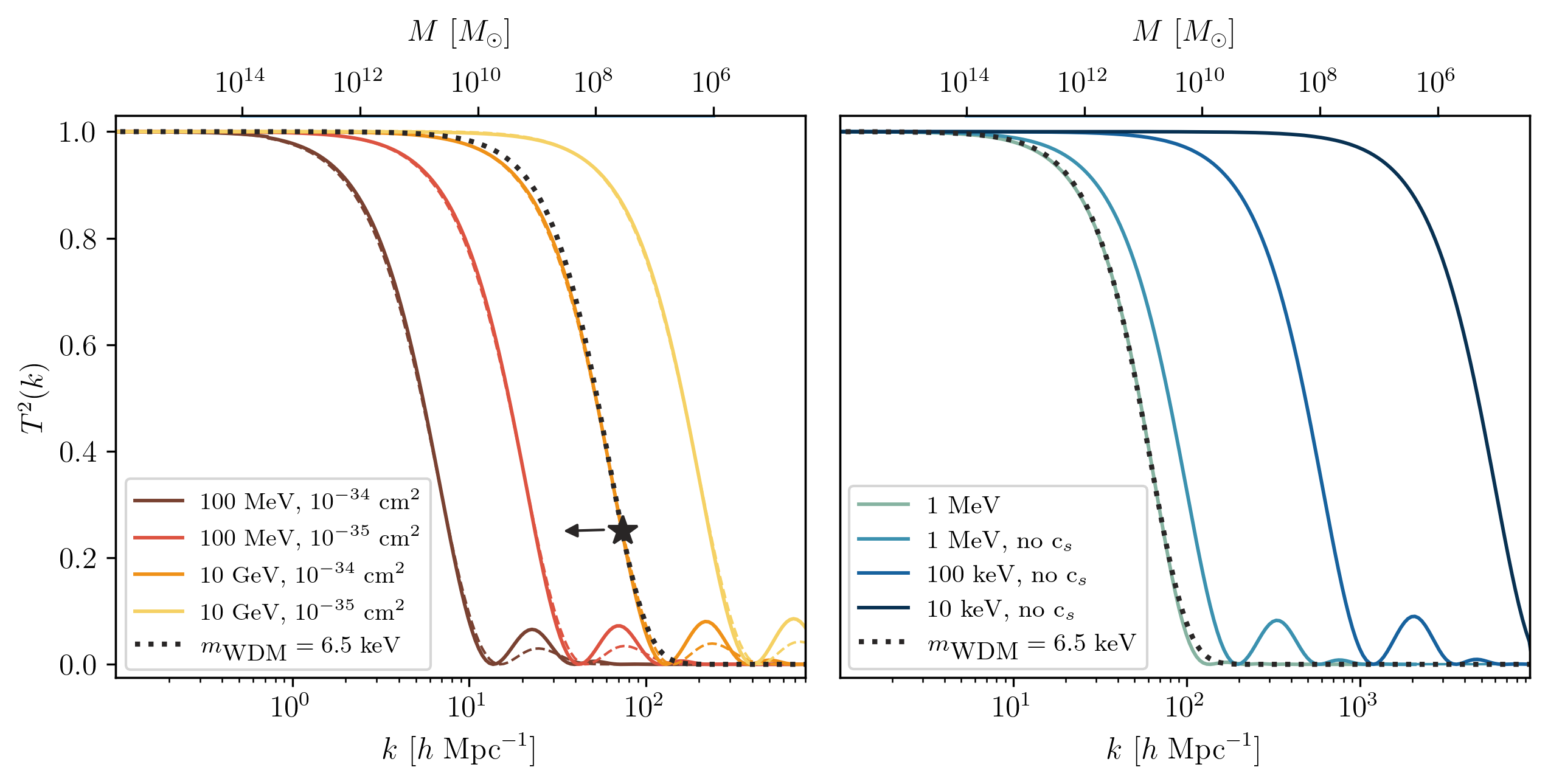}
    \caption{In the left panel, we illustrate the half-mode matching method
used to constrain momentum transfer cross section for a range of DM masses, as discussed in Section \ref{sec:approach}. We plot linear matter power
spectra ($P(k)$) normalized to CDM for both DM--photon scattering (solid) and DM--neutrino scattering (colored dashed) models. Cross sections denoted in the legend are for $\gamma$--DM models; for overlapping $\nu$--DM models, see Eq.~\ref{SigmaRatio}. We also plot the thermal-relic WDM model ($m_{\mathrm{WDM}}=6.5\textrm{keV}$) ruled out at $95\%$ confidence by the MW satellite population. DM--radiation scattering models that are more suppressed than this limit (i.e., with half-mode wavenumbers to left of the star) are disfavored by the MW satellite population. In the right panel, we illustrate the impact on $P(k)$ when accounting for DM sound speed in the Boltzmann hierarchy. The added suppression of power due to sound speed (i.e., the leftward shift of the transfer function) requires a smaller cross section to match observation.}
\label{fig:HalfModeExample}
\end{figure*}

When modeling DM interactions, additional collision terms must be added into the usual Boltzmann hierarchy; for a detailed treatment of the hierarchy in the presence of DM--radiation scattering, see Ref.~\cite{Becker_2021}. Here we highlight the primary modifications, where $i$ is replaced by $\gamma$ or $\nu$, depending on the scenario. The relevant Boltzmann equations read as follows:
\begin{equation}\label{boltzmann_DM}
    \begin{split}
        \delta_{i}^\prime = &-\frac{4}{3}\theta_{i} + 4\phi^\prime,
        \\
        \theta_{i}^\prime = & k^2\left(\frac{1}{4}\delta_i - \sigma_i\right) + k^2\psi - \Gamma_{i\text{-}b}\left(\theta_{i}-\theta_{b}\right) - \Gamma_{i\text{-}\chi}\left(\theta_{i}-\theta_{\chi}\right),
        \\
        \delta_{\chi}^\prime = &-\theta_{\chi} + 3\phi^\prime,
        \\
        \theta_\chi^\prime = &H\theta_{\chi}+c^{2}_{\chi}k^{2}\delta_{\chi} + k^2\psi -\Gamma_{\chi\text{-}i}\left(\theta_{\chi}-\theta_{i}\right).
    \end{split}
\end{equation} Here, $\delta_\chi$ ($\delta_i$) is the DM (radiation) density fluctuation, $\theta_\chi$ ($\theta_i$) is the DM (radiation) velocity divergence, $\psi$ and $\phi$ are scalar metric perturbations, $\sigma_i$ is the shear stress potential, $H$ is the Hubble rate, $c_\chi$ is the sound speed of the DM fluid, $\Gamma$ is the momentum exchange rate of the subscripted species, and primes denote derivative with respect to conformal time. Specifically, for interactions between DM and radiation, the momentum exchange rate is defined as
\begin{equation}\label{Gamma}
    \begin{split}
        \Gamma_{\chi\text{-}i} = \frac{4\rho_i}{3\rho_\chi}a\sigma_{\chi\text{-}i}n_\chi c = \frac{4}{3}\rho_i\frac{\sigma_{\chi\text{-}i}}{m_\chi}ac,
    \end{split}
\end{equation} 
where $\rho_\chi$ ($\rho_i$) is the DM (radiation) energy density, $a$ is the scale factor,  $n_\chi$ is the DM number density, and $c$ is the speed of light. Eq.~\ref{Gamma} provides the explicit dependence of momentum exchange on parameters $\sigma_{\chi\text{-}i}$ and $m_\chi$, which are commonly combined into the single dimensionless parameter
\begin{equation}\label{u}
    \begin{split}
    u_{\chi\text{-}i}=\frac{\sigma_{\chi\text{-}i}}{\sigma_{T}}\left(\frac{m_\chi}{100~\mathrm{GeV}}\right)^{-1},
    \end{split}
\end{equation} 
where $\sigma_{T}$ is the Thomson scattering cross section ($6.65\times10^{-25}$ $\textrm{cm}^2$).

As seen in Eq.~\ref{boltzmann_DM}, when coupled to radiation, DM has a nonzero sound speed $c_\chi$, given by \begin{equation}\label{sound speed}
    \begin{split}
    c^2_\chi=\frac{k_BT_{\chi}}{m_\chi}\left(1-\frac{1}{3}\frac{\partial lnT_{\chi}}{\partial lna}\right),
    \end{split}
\end{equation} where $k_B$ is Boltzmann's constant and $T_\chi$ is the temperature of the DM fluid. This non-relativistic treatment of $c_\chi$ is sufficient for our purposes, as interactions which suppress structure on MW satellite scales occur when $c_\chi$ is less than $\sim$$10\%$ the speed of light. The impact of interactions on $T_\chi$ must also be accounted for, and is given by
\begin{equation}\label{Temp}
    \begin{split}
    T_\chi^\prime=-2 H T_{\chi} - 2 \Gamma_{\chi\text{-}i} \left(T_{\chi} - T_{i} \right).
    \end{split}
\end{equation} 
Following previous literature, we neglect alterations in $T_i$ arising from DM scattering, and we model the temperature evolution of the radiation as in the standard CDM cosmology, $T_i=T_{i,0}(1+z)$. In the case of scattering with photons, $T_{i,0}$ is the CMB temperature today ($2.73$ K); for neutrino scattering, the radiation temperature acquires the usual factor of $(4/11)^\frac{1}{3}$ and reads as $1.95$ K.

For both DM--photon and DM--neutrino scattering, the momentum transfer cross section can be parametrized as a power law in the temperature of the DM fluid, where $\sigma_{\chi\text{-}i}$=$\sigma_0 T_\chi^n$ for $n \in \{-4,-2,0,2,4,6\}$); this parametrization captures a variety of microphysical interaction scenarios \cite{B_hm_2001,Boddy_2018,Becker_2021}. In this work, we only consider $n=0$---i.e., temperature independent interactions---and save additional values of $n$ for future work. 

Density perturbations are suppressed via two mechanisms in cosmologies in which DM is coupled to radiation. For high $m_\chi$, collisional damping dominates the suppression, and is captured by the drag term $\Gamma_{\chi\text{-}\gamma}(\theta_\chi-\theta_\gamma)$ in Eq.~\ref{boltzmann_DM}. For $m_\chi$ below $1$ MeV, the damping of acoustic peaks is primarily due to sound speed; in this case, damping is governed by the pressure term $c_\chi^2 k^2\delta_\chi$ in Eq.~\ref{boltzmann_DM}. For a detailed analysis of the effects of DM sound speed on the linear matter power spectrum, see Ref.~\cite{Stadler_2018}.

Linear matter perturbations corresponding to the halos that host the faintest MW satellites (i.e., subhalos of $\sim$$10^8$ solar masses) enter the cosmological horizon around redshift $10^6$, for cosmologies along our bound.\footnote{To arrive at this estimate for redshift, we convert $10^8$ $M_\odot$ to its critical length scale (see Eq.\ 5 of Ref.~ \cite{Nadler_2019_b}), then set this equal to the horizon size, solving for redshift.} For allowed models, kinetic decoupling of DM from radiation is bound to occur around or before this redshift, in order to avoid impact on the observed satellite population. Therefore, this critical redshift of $10^6$ can also be recovered from Eq.~\ref{Gamma}, by finding when the momentum transfer rate falls below the Hubble rate. Note that kinetic decoupling occurs well before matter--radiation equality, ensuring that scattering primarily affects the initial conditions for structure formation, rather than its growth and dynamics at later times. This allows the mapping of IDM models to the linear $P(k)$ of WDM.

\section{Method}
\label{sec:approach}

An interaction cross section $\sigma_{\chi\text{--}i}$ that leads to excessive suppression of the formation of halos known to host MW satellite galaxies is inconsistent with observations. We quantify the allowed level of suppression using the linear matter power spectrum, $P(k)$, of the $6.5~\mathrm{keV}$ thermal-relic WDM model ruled out by the MW satellite population at $95\%$ confidence \cite{Census_III}. This limit is based on a likelihood analysis of the MW satellite population observed by DES and PS1, accounting for observational incompleteness and selection effects \cite{Mao_2015,Drlica_Wagner_2020}.  

In WDM-like scenarios, the reduction in abundance of small halos and their associated dwarf galaxies results strictly from a small-scale suppression of linear power arising from non-negligible free streaming of DM particles. While the precise mechanism underlying the suppression differs, the $P(k)$ cutoff for thermal-relic WDM is similar to that observed in radiation scattering scenarios, and our approach leverages this similarity. That is, we match $P(k)$ for each scattering scenario to the ruled-out WDM model, thus mapping WDM bounds to bounds on $\sigma_{\chi\text{--}i}$. We note that the only difference between WDM and DM--radiation scattering transfer functions is the presence of low-amplitude dark acoustic oscillations (DAOs) at $k>100~h~{\textrm{Mpc}^{-1}}$ in the radiation case; see Fig.~\ref{fig:HalfModeExample}. However, these features have insufficient power to impact the subhalo mass function on the scales relevant to our analysis \cite{B_hm_2014,Schewtschenko_2015}. We therefore ignore DAOs in our mapping procedure.

\begin{table*}[t]
    \centering
    \begin{tblr}{|c|c|c|c|c|}
    \hline
    $m_\chi$ [GeV] & $\sigma_{\chi\text{-}\gamma}~[\mathrm{cm}^2]$  & $u_{\chi\text{-}\gamma}$ & 
    $\sigma_{\chi\text{-}\nu}~[\mathrm{cm}^2]$ &
    $u_{\chi\text{-}\nu}$ 
    \\
    \hline \hline
    $10^{-5}$ & $5.81\times10^{-44}$ & $8.74\times10^{-13}$ &
    \text{--} &
    \text{--}  \\ 
    \hline
    $10^{-4}$ & $5.85\times10^{-41}$ &  $8.79\times10^{-11}$ & $8.98\times10^{-41}$ & $1.35\times10^{-10}$  \\ 
    \hline
    $10^{-3}$ & $1.98\times10^{-38}$ &  
    $2.97\times10^{-9}$ & $3.16\times10^{-38}$  & $4.75\times10^{-9}$  \\
    \hline
    $10^{-2}$ & $3.88\times10^{-37}$ &  
    $5.83\times10^{-9}$ & $5.76\times10^{-37}$  & 
    $8.66 \times10^{-9}$ \\
    \hline
    $10^{-1}$ & $4.14\times10^{-36}$ &  
    $6.22\times10^{-9}$ & $6.11\times10^{-36}$  & $9.18\times10^{-9}$ \\
    \hline
    $10^{0}$ & $4.16\times10^{-35}$ &  
    $6.26\times10^{-9}$ & $6.15\times10^{-35}$  & $9.24\times10^{-9}$ \\
    \hline
    $10^{1}$ & $4.16\times10^{-34}$ &  
    $6.26\times10^{-9}$ & $6.15\times10^{-34}$  & $9.24\times10^{-9}$ \\
    \hline
    $10^{2}$ & $4.16\times10^{-33}$ &  
    $6.26\times10^{-9}$ & $6.15\times10^{-33}$  & $9.24\times10^{-9}$ \\
    \hline 
    \end{tblr}
    
    \caption{Upper bounds on temperature independent momentum-transfer cross section $\sigma_{\chi\text{-}i}$ and the related dimensionless interaction parameter $u_{\chi\text{--}i}=(\sigma_{\chi\text{--}i}/\sigma_T)(m_\chi/100~\mathrm{GeV})^{-1}$, taken as a function of dark matter mass $m_\chi$. These bounds were obtained using half-mode matching to the $T^2(k)$ cutoff associated with the thermal relic WDM mass lower bound (6.5 keV), as inferred by Ref.~\cite{Census_III} using DES and PS1 MW satellite observations.}
    \label{tab:Results}
    
\end{table*}

We compute the linear matter power spectra using the Boltzmann solver CLASS \cite{Blas_2011}. For $\gamma$--DM interactions, we rely on a modified version from Ref.~\cite{Becker_2021}, which accounts for DM sound speed.\footnote{\href{https://github.com/lesgourg/class_public}{CLASS v. 3.2.0.}}  For the $\nu$--DM case, we have modified the version used in Ref.~\cite{Mosbech_2021}\footnote{\href{https://github.com/MarkMos/CLASS_nu-DM}{https://github.com/MarkMos/CLASS\_nu-DM}} by including DM sound speed and temperature evolution (via Eqs.~\ref{sound speed} and \ref{Temp}). The transfer function reads as:
\begin{equation}\label{Tk}
    \begin{split}
    T^2(k)_{j}=\frac{P(k)_{j}}{P(k)_{\textrm{CDM}}},
    \end{split}
\end{equation} where $j\in [\nu,\gamma]$ labels the interacting DM scenario. The transfer function corresponding to the ruled-out 6.5 keV WDM model satisfies $T^2(k_{\mathrm{hm}})=0.25$ at $k_{\mathrm{hm}}=73~\textrm{h~Mpc}^{-1}$, where $k_{\mathrm{hm}}$ is the half-mode wavenumber.\footnote{This half-mode wavenumber was arrived at using Eqs. 5 and 3 in Refs.~\cite{Nadler_2021} and \cite{nadler2024cozmicicosmologicalzoomin}, respectively.}

For $\gamma$-DM scattering we consider a set of benchmark DM masses, $m_\chi\in[10\textrm{keV},100\textrm{GeV}]$. For each $m_\chi$, we find the $\sigma_{\chi\text{--}\gamma}$ that leads to a $T^2(k)_{\chi\text{--}\gamma}$ equaling $0.25$ at $k_\textrm{hm}$. Cross sections larger than these would produce a stronger suppression of power and are inconsistent with the observed MW satellite population at $95\%$ confidence. Smaller cross sections, which would result in larger $k_\textrm{hm}$, are allowed because the data do not yet place a lower limit on power at smaller scales. Thus, we obtain upper limits on $\sigma_{\chi\text{--}\gamma}$ as a function of $m_\chi$, shown in Fig.~\ref{fig:HalfModeExample}. 

We compute the bounds on $\sigma_{\chi\text{--}\nu}$ using the same methods; to be conservative, we only consider DM masses down to $100$ keV (i.e., $m_\chi\in[100\textrm{keV},100\textrm{GeV}]$). We make this choice because power spectra output by CLASS rise significantly on small scales ($k\gtrsim 100~h~\mathrm{Mpc}^{-1}$) in the neutrino-scattering $10$ keV case. Since this may be a numerical artifact, we defer further exploration of this regime to future work.

\section{Results}\label{sec:results}

\begin{figure*}[t]
    \centering
    \includegraphics[width=\linewidth]{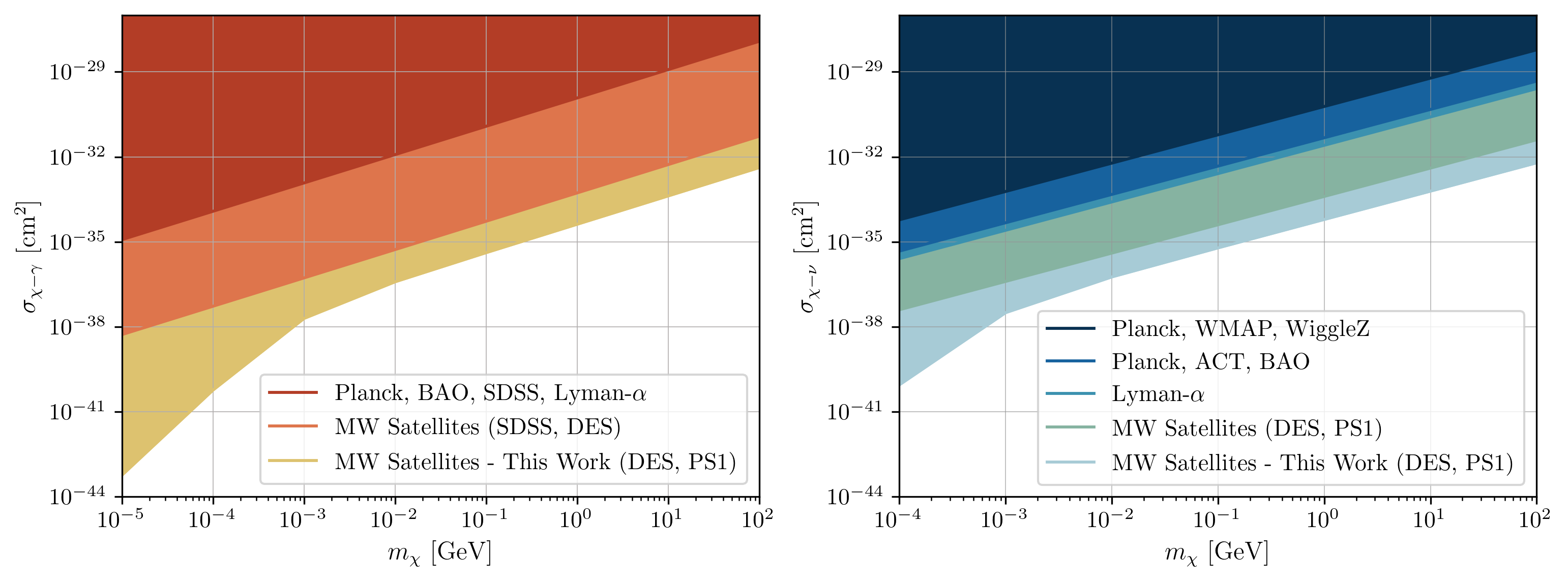}
    \caption{Upper bounds on temperature--independent momentum-transfer cross section are shown as a function of dark matter mass $m_\chi$, for DM scattering with photons (on left, down to $10$ keV) and neutrinos (on right, down to $100$ keV). Cross sections above the bottom-most gold (left) and light blue (right) boundaries are excluded at $95\%$ confidence by Ref.~\cite{Census_III}'s MW satellite population analysis. For comparison, we overplot $95\%$ confidence upper limits from Planck+BAO+SDSS+Lyman-$\alpha$ \cite{Becker_2021} and SDSS+DES MW satellites \cite{Escudero_2018} (for $\gamma$--DM), and from Planck+WMAP+WiggleZ \cite{Escudero_2015}, Planck+ACT+BAO \cite{giarè2023hints}, Lyman-$\alpha$ forest \cite{Hooper_2022}, and DES+PS1 MW satellites \cite{Akita_2023} (for $\nu$--DM). Our results improve upon existing bounds on cross section by 1 to 5 orders of magnitude.}
\label{fig:ParamSpaceDouble}
\end{figure*}

In Table~\ref{tab:Results}, we present our $95\%$ confidence level bounds on interaction cross sections $\sigma_{\chi\text{-}\gamma}$ and $\sigma_{\chi\text{-}\nu}$ for a range of $m_\chi$. We also report the limits on the dimensionless parameters $u_{\chi\text{--}\gamma}$ and $u_{\chi\text{--}\nu}$ (interchangeable with $\sigma_{\chi\text{-}i}$ via Eq.~\ref{u}). Finally, we plot our $\sigma_{\chi\text{-}\gamma}$ and $\sigma_{\chi\text{-}\nu}$ bounds in Fig.~\ref{fig:ParamSpaceDouble}, where the shaded regions of the parameter space are excluded at the $95\%$ confidence level. We show a comparison with previous results in the same figure. 

For $\gamma$--DM scattering, previous bounds were derived from linear cosmological probes (combining the Planck 2018 temperature, polarization, and lensing anisotropy with BAO, SDSS, BOSS, and Lyman-$\alpha$ forest data) in Ref.~\cite{Becker_2021} and from the MW satellite population (comparing total predicted subhalo counts with completeness-corrected SDSS and DES data) in Ref.~\cite{Escudero_2018}. For $\nu$--DM scattering, previous bounds were derived from Planck 2013 temperature anisotropy in combination with the Wilkinson Anisotropy Microwave Probe (WMAP) polarization and WiggleZ full-shape galaxy power spectrum in Ref.~\cite{Escudero_2015}; from Planck 2018 temperature, polarization, and lensing anisotropy in combination with ACT DR4, and BAO (6dfGS, SDSS, BOSS DR12) in Ref.~\cite{giarè2023hints}; from Lyman-$\alpha$ in Ref.~\cite{Hooper_2022}; and from the MW satellite population (comparing semi-analytic predictions of subhalo counts with completeness-corrected DES and PS1 data) in Ref.~\cite{Akita_2023}. For both $\gamma$--DM and $\nu$--DM scenarios, Lyman-$\alpha$ bounds were derived using HIRES/MIKE spectra probing wavenumbers up to $10~h~\mathrm{Mpc}^{-1}$~\cite{Archidiacono_2019}.

Overall, we find that our results strengthen previous bounds on the DM--radiation interaction cross section derived using MW satellites by 1 order of magnitude at high DM masses, for both interaction scenarios. Our inclusion of DM sound speed further strengthens the bounds for $m_\chi\lesssim1$ MeV; this improvement reaches an additional 2 orders of magnitude at a DM mass of $100$ keV. At $10$ keV, this improvement grows to 5 orders of magnitude for the $\gamma$--DM scenario.

For both scenarios, the mass dependence of our bounds is linear for $m_\chi\gtrsim1$ MeV but approaches $m_\chi^3$ at lower masses. This behavior is due to the two distinct physical mechanisms that suppress structure in DM--radiation interaction scenarios; namely,  collisional damping and sound waves in the DM fluid. Each necessitates the inclusion of a term in the DM velocity divergence equation, Eq.~\ref{boltzmann_DM}, and the different $m_\chi$ scalings of these terms drives each to dominate in a different mass regime. For $m_\chi\lesssim1$ MeV, the drag term ($\Gamma_{\chi\text{-}i}(\theta_\chi-\theta_\gamma)$) becomes negligible compared to the pressure term ($c^{2}_{\chi}k^{2}\delta_{\chi}$), and the opposite is true for $m_\chi\gtrsim1$ MeV. In order to ensure---for all $m_\chi$---a suppression of linear power matching the WDM model ruled out by Ref.~\cite{Census_III}, $c_\chi$ must be constant along our bound at low masses, and likewise $\Gamma_{\chi\text{-}i}$ for high (i.e., at redshifts that determine $P(k)$ suppression on MW satellite scales, $\textrm{z}\sim$$10^6$). These conditions, taken with Eq.~\ref{sound speed}, a temperature scaling of $T_\chi \sim \sigma_{\chi\text{--}i}/m_\chi^2$, and Eq.~\ref{Gamma}, give the observed mass dependencies. The mass scale of the transition---occurring here at $m_\chi\sim1$ MeV---results from the half-mode scale of the ruled-out WDM model. The transition would move to higher mass should lower interaction strength be required to match observation. 

Comparing the two scenarios themselves, we find that the ratio of the $\gamma$--DM and $\nu$--DM bounds is approximately $0.69$ at high DM masses. This is expected, as the sound speed term is negligible in this regime, thus requiring $\Gamma_{\chi\text{-}\gamma}=\Gamma_{\chi\text{-}\nu}$ at the redshifts when interactions are efficient to produce the same half-mode scale. From this fact and Eq.~\ref{Gamma}, we find that
\begin{equation}\label{SigmaRatio}
    \frac{\sigma_{\chi\text{-}\gamma}}{\sigma_{\chi\text{-}\nu}}=\frac{\rho_\nu}{\rho_\gamma}=\frac{7}{8}N_{\mathrm{eff}}\times\left(\frac{4}{11}\right)^{4/3}\approx 0.69.
\end{equation} 
At low DM masses, the ratio of bounds is less trivial, as the degree of suppression becomes sensitive to the temperature of the scattering species (via Eqs.~\ref{sound speed} and \ref{Temp}) as well as its density. 

Additionally, we find a difference in DAO structure between the two scenarios, as observed in Fig.~\ref{fig:HalfModeExample}. Since interactions critical to suppression on MW satellite scales occur well before recombination, this could be due to the additional baryon coupling in the photon case as opposed to the neutrino. Interestingly, despite the shift in half-mode scale of the $P(k)$ cutoff, the shape of suppression is similar whether the effects of sound speed are included or not. In particular, only a mild difference in DAO height is evident (see the right panel of Fig.~\ref{fig:HalfModeExample}). It will be interesting to analytically model this similarity, as well as the size of the transfer function shift, in future work.  

Lastly, the DM--radiation interactions that suppress structure on MW satellite scales occur at $z\gtrsim 10^6$, when neutrinos are still relativistic. We have thus used a common approximation in the literature and treated neutrinos as massless. However, a preliminary investigation into the massive neutrino scenario for high DM mass indicates an approximately $50\%$ weaker bound over the massless case. This reduction is a small effect relative to the order of magnitude difference we obtain relative to previous constraints in the literature, and is consistent with the non-relativistic treatment of momentum transfer between DM and massive neutrinos. We save a complete investigation of this scenario for future work.

\section{Discussion and Conclusions}
\label{sec:conclusion}

Interactions between DM and SM radiation in the early Universe suppress small-scale matter perturbations, ultimately reducing the abundance of low-mass halos and the dwarf galaxies they host. We used the MW satellite population to place new constraints on DM--radiation scattering. We did so by leveraging the lower limit on thermal-relic WDM mass that was derived using DES and PS1 data, i.e., $m_\textrm{WDM}=6.5$ keV. We related the suppression of the linear matter power spectrum for this ruled-out model to DM--radiation interaction scenarios, producing upper bounds on the associated interaction cross sections as a function of DM mass (see Tab.~\ref{tab:Results} and Fig.~\ref{fig:ParamSpaceDouble}). At large DM masses, the new bounds are 1 order of magnitude more stringent than previous studies. We extended beyond existing analyses by modeling the effects of DM sound speed on the linear matter power spectrum. This increased the constraining power of the current data for DM masses below $\sim1~\mathrm{MeV}$, eliminating many additional orders of magnitude in cross section.

For both DM--neutrino and DM--photon scattering, the improvement of our bounds over previous MW satellite analyses ~\cite{B_hm_2014,Escudero_2018, Akita_2023} is consistent with the corresponding improvement in WDM constraints derived in Ref.~\cite{Census_III}. This improvement results from detailed forward-modeling of the DES and PS1 satellite population in Ref.~\cite{Census_III}, which uses the latest observational selection functions from Ref.~\cite{Drlica_Wagner_2020} and models satellites associated with the Large Magellanic Cloud; see Ref.~\cite{Nadler:2021dft} for a detailed discussion. We note that alternative WDM limits derived from recent MW satellite analyses \cite{Newton:2020cog,Dekker:2021scf} translate to weaker bounds on DM--radiation models. For example, adopting the $3.99~\mathrm{keV}$ limit from Ref.~\cite{Newton:2020cog} weakens our constraints by a factor of $\sim$$3$; however, this is a small change relative to the orders-of-magnitude improvement we derive. In future work, it will be important to compare different MW satellite analyses in detail to understand such differences~\cite{Nadler:2021dft,Newton:2024jsy}.

Since the effects of DM--radiation scattering at a given cross section are generally more apparent on smaller scales, our bounds are stronger than those derived from the CMB anisotropy and other linear cosmological probes; see Fig.~\ref{fig:ParamSpaceDouble}. However, our results are in tension with recent reports of a mild preference---at $68\%$ confidence---for non-vanishing $\nu$--DM interactions in both the Lyman-$\alpha$ forest and CMB, the latter when including very small angular scales \cite{giarè2023hints,Hooper_2022,Brax_2023,Brax_2023_1}. According to our mapping, the model preferred by Ref.~\cite{giarè2023hints} would suppress the linear matter power spectrum down to wavenumbers of $k\sim 1~h~\mathrm{Mpc}^{-1}$, suppressing halo formation up to $\sim 10^{13}~M_{\mathrm{\odot}}$. Future dedicated comparison studies may be needed to understand the source of this tension. 

Finally, comparison with bounds set by high-energy neutrino probes---such as Ice Cube---require modeling of interactions at a broad range of neutrino energies \cite{Cline_2023,Cline_2023_b, heston2024constraining}. The results presented here, however, only probe neutrino scattering at energies on the order of $0.1$ to $1$ keV---as set by the horizon entry of the perturbation modes that source MW substructure. Future targeted comparisons between probes will be informative. Likewise, scenarios in which a sub-component of DM scatters with radiation---and the remainder behaves like CDM---provide a rich territory for future investigation. A degeneracy naturally arises in such scenarios, between the fraction of interacting DM and its scattering cross section; dedicated analysis will be needed to understand the consequences on small-scale structure.

Looking to the future, a coordinated effort using a variety of cosmological probes---potentially in combination with direct detection experiments---will be required to look for signatures of DM--radiation scattering. Promising probes include CMB anisotropies at small angular scales \cite{SimonsObservatory:2018koc,CMB-S4:2022ght}, high-energy astrophysical events, Lyman-$\alpha$ forest power spectra, 21-cm cosmology \cite{Dey_2023,Mosbech_2023}, gravitational-wave event rates \cite{Mosbech:2022nkk}, galaxy surveys \cite{Escudero_2015,LSSTDarkMatterGroup:2019mwo,nadler2024forecasts}, and other probes \cite{gluscevic2019cosmological}. In particular, next generation observational facilities, including the Vera C.\ Rubin Observatory and the Nancy Grace Roman Space Telescope, will detect faint dwarf galaxies throughout the Local Volume and beyond, probing progressively smaller interaction cross sections---not just for DM-radiation scattering, but for all scenarios that suppress structure on small scales~\cite{nadler2024forecasts}. 

In addition to new data, theoretical developments in galaxy--halo connection modeling---as well as the identification of novel observable probes of DM--radiation interactions, including subhalo density profiles and signatures from late-time dynamical effects \cite{Schewtschenko_2015,Schewtschenko_2016,Molin__2019,bramante2024illuminatingdarkobjectsdark}---will further advance present analyses. In parallel, the development of self-consistent cosmological simulations that can forward model the MW satellite population in the presence of DM--radiation scattering will be critical to move analyses beyond constraints and into a regime where new physics can be robustly identified, using near-field cosmology and beyond.

\section*{Acknowledgments}
We thank Kim Boddy for her helpful comments on the manuscript. W.C. acknowledges Trey Driskell for his valuable feedback and explanations, as well as Yacine Ali-Haïmoud for his insights; W.C. also thanks Markus Mosbech for his time responding to software related inquiries early on in this project.
V.G. acknowledges the support from NASA
through the Astrophysics Theory Program, Award Number 21-ATP21-0135, the National Science Foundation (NSF) CAREER Grant No. PHY-2239205, and from the Research Corporation for Science Advancement under the Cottrell Scholar Program. This research was supported in part by Grant NSF PHY-2309135 to the Kavli Institute for Theoretical Physics (KITP).

\bibliography{main}

\end{document}